\documentclass{desyproc}

\begin{document}

\title{Axion-like Particles from String Compactifications}

\author{{\slshape Michele Cicoli${}^{1,2,3}$}\\[1ex]
$^1$ Dipartimento di Fisica ed Astronomia, Universit\`a di Bologna, Bologna, Italy. \\
$^2$ INFN, Sezione di Bologna, Italy.\\
$^3$ Adbus Salam ICTP, Strada Costiera 11, Trieste 34014, Italy.}

\contribID{Cicoli\_Michele}

\desyproc{DESY-PROC-2013-04}
\acronym{Patras 2013}

\maketitle

\begin{abstract}
We review how axion-like particles (ALPs) naturally emerge
in the low-energy effective field theory of string compactifications.
We focus on the study of their mass spectrum and couplings, stressing
that they depend on the mechanism used to fix the moduli.
We present concrete examples where either open or closed string modes
behave as QCD axions which do not overproduce cold dark matter.
Relativistic ALPs can also be produced by the decay of the lightest modulus which drives reheating.
These ALPs contribute to dark radiation and could be
detected via axion-photon conversion in astrophysical magnetic fields.
\end{abstract}

\section{Axions and strings}

\subsection{Axions as probes of high energy physics}

The QCD axion $a_{\rm QCD}$ is the most plausible explanation of the strong CP problem.
Its mass and couplings to ordinary particles are set by its decay constant $f_{a_{\rm QCD}}$
which must lie in the phenomenologically allowed window $10^9\,{\rm GeV} \lesssim f_{a_{\rm QCD}} \lesssim 10^{12}\,{\rm GeV}$,
where the lower bound comes from the non-observation of cooling of stars due to axion emission,
while the upper bound is due to the overproduction of axionic cold dark matter (DM).

Note that the upper bound is somewhat looser since it assumes a
standard post-inflationary cosmological evolution which does not apply to
cases where ordinary particles are diluted by the decay of new gravitationally coupled scalars.
In particular, axionic DM can be diluted if this decay leads to a reheating temperate below
the QCD phase transition, $T_{\rm rh} < \Lambda_{\rm QCD}\simeq 200$ MeV.
The case of maximum dilution is obtained when $T_{\rm rh}$ is just above Big Bang Nucleosynthesis (BBN),
$T_{\rm rh} \gtrsim T_{\rm BBN}\simeq 3$ MeV,
which raises the upper bound to $f_{a_{\rm QCD}} \lesssim 10^{14}$ GeV \cite{Fox:2004kb}.
Larger values of $f_{a_{\rm QCD}}$ require some tuning of the initial misalignment angle.
Moreover, the transparency of the universe for TeV gamma-rays \cite{Transparency}
and the anomalous cooling of white dwarfs \cite{Isern:2012ef}
point together to a very light ALP with an intermediate scale decay constant.

All these constraints reveal that axions are associated with a very high energy scale.
Hence, it is natural to search for them in beyond the Standard Model (BSM) theories like string theory.
In fact, the low-energy limit of string compactifications
yields an effective field theory (EFT) with promising QCD axion candidates~\cite{StringyAxions,Cicoli:2012sz},
or even an `axiverse' containing a plethora of light ALPs
with a logarithmically hierarchical mass spectrum~\cite{Arvanitaki:2009fg}.
The strongest constraint on the axiverse comes
from the production of isocurvature fluctuations during inflation.
Their fraction with respect to the total amplitude of adiabatic plus isocurvature fluctuations
is $\beta_{\rm iso}<0.039$ at 95\% CL \cite{Ade:2013uln}.
If all DM consists of axions, $\beta_{\rm iso}$ is set by the inflationary scale $H_{\rm inf}$, the
axion decay constants $f_{a_i}$ and the initial misalignment angles $\theta_i$:
$\beta_{\rm iso} \simeq 4\cdot 10^7 \sum_i \left(\frac{H_{\rm inf}}{\theta_i f_{a_i}}\right)^2\lesssim 4\cdot 10^{-2}$.
Considering $n_a$ axions with $f_{a_i}=f$ and $\theta_i=\theta$ $\forall i=1,...,n_a$, the previous constraint becomes
$f\gtrsim \sqrt{n_a}/\left(3\theta\right) 10^5\,H_{\rm inf}$.
The inflationary scale sets also the amplitude of the tensor modes.
In particular, a detection of gravitational waves by the Planck satellite would imply
$H_{\rm inf}\simeq M_{\rm GUT}\simeq 10^{16}$ GeV, which, in turn, would
rule out the axiverse since it would require $f\gtrsim 10^{21}$ GeV
(for $n_a\sim\mathcal{O}(100)$ and $\theta\sim\mathcal{O}(\pi)$).
Notice that such a high inflationary scale would be a problem also for cases with just one
light axion, the QCD axion, if it contributes to DM and its initial misalignment angle is not
tuned to small values.

Due to this interesting possibility to put stringy ideas to experimental test,
it is crucial to give a solid answer to each of the following questions:
\begin{enumerate}
\item What kind of ALP masses and couplings should we expect from string compactifications?
Is it generic to obtain an axiverse?

\item Can we build concrete examples of globally consistent semi-realistic chiral models
with stabilised moduli and an explicit QCD axion candidate?

\item What can be the r\^ole played by additional ultra-light axions? How can we detect them?
\end{enumerate}
This last question is particulary important since
string theory naturally provides particles which can behave as the QCD axion,
even if its presence might be considered as required only by the solution of the strong CP problem,
and so as a feature of BSM theories which have no relation to string theory. On the other hand, ultra-light ALPs
do not play any r\^ole in the solution of the strong CP problem. Hence,
they can be considered as truly stringy predictions since their presence
in BSM theories does not seem to be needed for any fundamental purpose.

\subsection{Axions from string compactifications}

The massless spectrum of any string theory contains antisymmetric forms whose
Kaluza-Klein reduction gives rise to ALPs in the low-energy 4D theory.
These axions are closed strings living in the bulk which come along with shift symmetries inherited
from higher-dimensional gauge symmetries. They are the imaginary part $a$ of a complex scalar field
$T=\tau+{\rm i} a$, where $\tau$ is the `saxion' field. This is a modulus whose
vacuum expectation value (VEV) determines the size of the extra dimensions
and key-features of the EFT like gauge and Yukawa couplings.
The saxion $\tau$, if long-lived, can cause a cosmological moduli problem (CMP).
In fact, when $H\sim m_\tau$, $\tau$ starts oscillating around its minimum and stores energy.
Given that it redshifts as matter, it quickly comes to dominate the energy density of the universe.
When $\tau$ decays at $H \sim \Gamma \sim \epsilon^2 m_\tau$ where $\epsilon \equiv \frac{m_\tau}{M_P}\ll 1$,
it reheats the universe to a temperature of order $T_{\rm rh} \sim \epsilon^{1/2} m_\tau$.
Requiring $T_{\rm rh} >T_{\rm BBN}$, one obtains a strong lower bound
on the modulus mass: $m_\tau \gtrsim \mathcal{O}(50)$ TeV.

The number $n_a$ of these ALPs depends on the topology of the extra dimensions
and for a generic Calabi-Yau (CY) one has $n_a \sim \mathcal{O}(100)$.
In type II theories, some of these axions are removed from the low-energy spectrum by the
orientifold projection which breaks the $\mathcal{N}=2$ 4D theory down to a chiral
$\mathcal{N}=1$ theory. However, this operation does not significantly change the order of magnitude of
the number of closed string axions left over.

Axions also arise as open strings living on space-time filling branes
which wrap some of the extra dimensions and support visible or hidden gauge theories.
These ALPs are phases $\psi_a$ of matter fields $C = |C|\,e^{{\rm i}\psi_a}$ whose radial part
breaks an effective global Peccei-Quinn $U(1)$ symmetry by getting a non-zero VEV via D-term stabilisation.
The number of these open string axions depends on the details of the brane set-up and
it can also be rather large in cases with large numbers of branes (if allowed by tadpole cancellation).

The dynamics stabilising the moduli determines which of these axions can be kept light:
\begin{itemize}
\item \emph{D-term stabilisation}: In the presence of an anomalous $U(1)$,
one has a D-term scalar potential which schematically looks like
(assuming just one charged open string mode $C$):
\begin{equation}
V_D \sim g^2 \left(|C|^2-\xi\right)^2\,,
\end{equation}
where $\xi$ is the Fayet-Iliopoulos (FI) term that depends on
the closed string modulus $\tau$ charged under the anomalous $U(1)$. Setting
the D-terms to zero implies $|C|^2=\xi(\tau)$. In turn,
the gauge boson gets a St\"uckelberg mass by eating up the axion corresponding to the combination
of $|C|$ and $\tau$ fixed by the D-term condition:
\begin{equation}
M_{U(1)}^2 \sim g^2 \left[\left(f_a^{\rm open}\right)^2+\left(f_a^{\rm closed}\right)^2\right]\,,
\end{equation}
where the open and closed string axion decay constants $f_a^{\rm open}$ and $f_a^{\rm closed}$
are given by:
\begin{equation}
\left(f_a^{\rm open}\right)^2=\langle |C|^2 \rangle =\xi \simeq \left|\frac{\partial K}{\partial \tau}\right|
\qquad\text{and}\qquad \left(f_a^{\rm closed}\right)^2 \simeq \frac{\partial^2 K}{\partial \tau^2}\,.
\label{fs}
\end{equation}
Here $K$ is the K\"ahler potential of the 4D $\mathcal{N}=1$ EFT.
If $f_a^{\rm open} \gg f_a^{\rm closed}$, the combination of moduli fixed by D-terms
is mostly $|C|$ and $\psi_a$ is eaten by the anomalous $U(1)$. If instead $f_a^{\rm open} \ll f_a^{\rm closed}$,
the modulus frozen by D-terms is $\tau$
and the axion eaten is $a$. Note that the $U(1)$ mass generated in this way is
in general of order the string scale.

\item \emph{F-term stabilisation}: The axion $a$ enjoys a shift symmetry $a \to a + {\rm const}$
which is broken only by non-perturbative effects. On the contrary, the saxion $\tau$ is not protected
by any symmetry, and so can develop a potential at both perturbative and non-perturbative level:
\begin{enumerate}
\item If $\tau$ is fixed by perturbative effects, then $a$ is exactly massless at this level
and its direction is lifted only by subleading non-perturbative effects. In this case $\tau$
and $a$ are stabilised by different effects, and so their masses can be different. In particular,
$\tau$ can satisfy the cosmological bound $m_\tau\gtrsim \mathcal{O}(50)$ TeV with $a$ almost massless.

\item If perturbative effects are made negligible by tuning some parameters,
both $\tau$ and $a$ are fixed at non-perturbative level. Hence they get a mass of the same order of magnitude,
rendering the axions too heavy: $m_a \sim m_\tau\gtrsim \mathcal{O}(50)$ TeV. These masses are generically of order the gravitino mass
$m_{3/2}$, and so if $m_a$ is lowered to smaller values relevant for phenomenology like $m_a \sim \mathcal{O}({\rm meV})$
(assuming a solution to the CMP), one would obtain a tiny scale of supersymmetry (SUSY) breaking.
\end{enumerate}
These considerations imply that very light axions can arise in the 4D EFT
only when some moduli are fixed perturbatively. Moreover, stringy instantons or
gaugino condensation should not develop a mass for the axions which is too large.
In the case of the QCD axion, these non-perturbative effects should not be larger than QCD instantons.
\end{itemize}

\subsection{Axions and chiral model building}

In order to find a viable QCD axion from string theory,
besides understanding how to keep the axions light, one should also embed QCD in
CY compactifications. More generally, one should build
consistent compact models with stabilised moduli and chiral non-Abelian gauge theories.

Type II theories seem to be a promising framework to achieve this goal because MSSM-like theories live on
localised objects called D-branes. Chiral model building becomes therefore a \emph{local} issue,
and so decouples from moduli stabilisation which is a \emph{global} issue.
This allows a separate study of the two problems with the idea of combining the two independent solutions.

Focusing on type IIB compactifications, semi-realistic chiral models can be built using:
\begin{enumerate}
\item Intersecting fluxed D7-branes wrapping cycles in the geometric regime;

\item Fractional D3-branes at CY singularities obtained by shrinking some cycles to zero size.
\end{enumerate}
The decoupling between chirality and moduli stabilisation
is actually only a leading order effect since various tensions arise
once chiral models are embedded in explicit CY constructions.
A stabilisation scheme which avoids all these tensions is the
LARGE Volume Scenario (LVS) \cite{LVS} which allows the construction of globally consistent compact models
where the visible sector can be either in
the geometric~\cite{Cicoli:2011qg} or in the singular~\cite{Singular} regime.

\subsection{Axions and moduli stabilisation}

In LVS models, the moduli are fixed by the interplay of all possible contributions
to the scalar potential: tree-level background fluxes, D-terms, $\alpha'$ and $g_s$ perturbative corrections,
and non-perturbative effects. This allows us to illustrate
the implications of any moduli fixing effect for the dynamics of the axion fields~\cite{Cicoli:2012sz}.
Let us summarise the LVS strategy to fix the moduli:
\begin{itemize}
\item The dilaton and complex structure moduli are fixed at semi-classical
level by turning on background fluxes. The VEV of the flux-generated
superpotential is naturally $W_0\sim\mathcal{O}(1)$.

\item The $h^{1,1}$ K\"ahler moduli $T_i=\tau_i + {\rm i}a_i$, where $\tau_i$
is the volume of the $i$-th internal 4-cycle and $a_i$ the corresponding axion,
are flat directions at tree-level due to the no-scale cancellation.

\item The scalar potential for the $T$-moduli can be expanded in inverse powers of the CY volume $\mathcal{V}$.
For $\mathcal{V}\gg 1$ (as required to trust the EFT), the dominant effect
comes from D-terms.

\item For vanishing open string VEVs, $d$ combinations of $T$-moduli are fixed by the D-term potential,
and so $d$ axions get eaten by anomalous $U(1)$s. If $d=h^{1,1}$,
the D-term conditions force the CY volume to collapse to zero size.
Thus one has to choose a brane set-up and fluxes such that $d< h^{1,1}$.
In this case, D-term fixing leaves $h^{1,1}-d \geq 1$ flat directions.

\item $n_{\rm np}$ del Pezzo (dP) divisors generate
single non-perturbative contributions to the superpotential whose existence
is guaranteed by the rigidity of these cycles and the absence of any chiral intersection
with the visible sector. Hence $n_{\rm np}$ K\"ahler moduli together with their corresponding axions
develop a mass of order $m_{3/2}$ due to non-perturbative effects.

\item The remaining $n_{\rm ax} = h^{1,1}-n_{\rm np}-d$ moduli tend to be fixed perturbatively
by $\alpha'$ or $g_s$ effects. Thus the corresponding axions remain massless and are good QCD axion candidates.
The main example is given by the volume mode $\mathcal{V}$ which develops an exponentially large VEV
due to $\alpha'$ corrections: $\mathcal{V} \sim W_0 \,e^{\frac{2\pi}{N g_s}}$ where $N$ is the rank
of an $SU(N)$ theory which undergoes gaugino condensation ($N=1$ for D3-instantons).
Another example is given by two intersecting local blow-up modes supporting the visible sector,
with one combination fixed by D-terms and the other by string loop corrections \cite{Cicoli:2012sz}.

\item The $n_{\rm ax}$ massless axions are lifted by higher-order instanton effects.
Given that for an arbitrary CY $h^{1,1}\sim\mathcal{O}(100)$,
$n_{\rm ax}$ might turn out to be very large giving rise to an axiverse.
\end{itemize}

\section{Axions in the LARGE Volume Scenario}

\subsection{Sequestered models}
\label{Seq}

Type IIB LVS models are particularly interesting also because
the moduli mass spectrum and couplings can be computed explicitly.
Consequently, one can study the post-inflationary cosmological evolutions of these models in detail.

The volume mode $\phi$ turns out to be the lightest modulus with a mass of order:
\begin{equation}
m_\phi\simeq m_{3/2}\sqrt{\epsilon}\ll m_{3/2} \qquad\text{where}\qquad\epsilon\equiv
\frac{m_{3/2}}{M_P}\simeq \frac{W_0}{\mathcal{V}} \simeq e^{-\frac{2 \pi}{N g_s}}\ll 1\,.
\end{equation}
Given that $\phi$ is lighter then the gravitino, there is automatically no cosmological problem
associated with a possible decay of $\phi$ into gravitini.
However, in gravity mediation one has in general $m_{3/2} \simeq \mathcal{O}(M_{\rm soft})$,
and so the requirement of TeV-scale SUSY implies $m_\phi \simeq \mathcal{O}(1)$ MeV.
Such a light modulus would definitely decay after BBN.

A viable solution to this cosmological problem relies on models with D3-branes at singularities.
The simplest version of these models has a CY volume $\mathcal{V}\,= \tau_b^{3/2}-\tau_{\rm np}^{3/2} - \tau_{\rm vs}^{3/2}$,
where $\tau_b$ is the `big' cycle controlling the overall volume ($\phi$ is the corresponding canonically normalised field),
$\tau_{\rm np}$ is a rigid divisor supporting non-perturbative effects and $\tau_{\rm vs}$ is the visible sector cycle.
$\tau_{\rm vs}$ collapses to zero size due to D-terms without breaking SUSY.
SUSY is instead broken by the $T$-moduli living in the bulk which develop a potential at $\alpha'$ and non-perturbative level:
\begin{equation}
V \sim \frac{\sqrt{\tau_{\rm np}}}{\mathcal{V}}\, e^{-\frac{4\pi  \tau_{\rm np}}{N}}
- W_0 \,\frac{\tau_{\rm np}}{\mathcal{V}^2}\,e^{-\frac{2\pi  \tau_{\rm np}}{N}}+\frac{W_0^2 \xi }{g_s^{3/2}\mathcal{V}^3}
\quad\Rightarrow\quad \langle\tau_{\rm np}\rangle \sim \frac{1}{g_s}>1\,,\quad\langle\mathcal{V}\rangle \sim \frac{W_0}{\epsilon}\gg 1\,.
\nonumber
\end{equation}
This set-up gives rise to \emph{sequestered} models with suppressed soft-terms \cite{sequester}:
\begin{equation}
M_{\rm soft}\simeq m_{3/2}\epsilon \ll m_\phi \simeq m_{3/2}\sqrt{\epsilon}\ll m_{3/2}\,.
\end{equation}
For $\epsilon \simeq \mathcal{O}(10^{-7})$, one obtains the following mass hierarchy:
\begin{equation}
M_{\rm soft} \simeq \mathcal{O}(1)\,{\rm TeV} \ll m_\phi
\simeq \mathcal{O}(5\cdot 10^6) \,{\rm GeV} \ll m_{3/2} \simeq \mathcal{O}(10^{11})\,{\rm GeV}\,, \nonumber
\end{equation}
which avoids any CMP, leads to low-energy SUSY and allows a high string scale,
$M_s \simeq M_P \sqrt{\epsilon}\simeq \mathcal{O}(10^{15})$ GeV, suitable for GUT and inflationary model building.
The axion $a_{\rm np}$ acquires a mass of order $m_{3/2}$, the volume axion $a_b$ remains in practice massless
since $m_{a_b} \sim M_P\, e^{-2\pi \mathcal{V}^{2/3}}\sim 0$, while the local axion $a_{\rm vs}$ gets eaten up by an anomalous $U(1)$.
This is always the case for arbitrary dP singularities where all
local closed string axions get eaten up by anomalous $U(1)$s.
However, some $a_{\rm vs}$ axions might remain light for more complicated singularities.
On top of these closed string axions, there could also be some open string ones whose properties are more model-dependent.
We shall discuss their r\^ole in Sec. \ref{SeqAx}.

\subsection{Reheating}

Reheating after the end of inflation is caused by the decay of the lightest modulus $\phi$
since it is the most long-lived. This decay injects entropy into the thermal bath diluting any
previous matter-antimatter asymmetry, axionic DM (if $T_{\rm rh}<\Lambda_{\rm QCD}$), and standard thermal LSP
DM (if $T_{\rm rh}<T_{\rm freeze-out}\simeq \frac{m_{\rm LSP}}{20}$). On the other hand,
the decay of $\phi$ can recreate non-thermally baryon asymmetry~\cite{Clado} and LSP DM~\cite{Bobby,Allahverdi:2013noa}, as well as relativistic
particles which behave as dark radiation (DR)~\cite{DR}. It is therefore crucial to study the
decay of the volume mode $\phi$ which takes place when:
\begin{equation}
H \simeq \Gamma_\phi = \frac{c}{2\pi} \frac{m_\phi^3}{M_P^2}
\qquad\Rightarrow\qquad T_{\rm rh} = c^{1/2} \left( \frac{m_\phi}{5\cdot 10^6\, {\rm GeV}}\right)^{3/2}\,
\mathcal{O}(1)\,{\rm GeV}\,,
\label{Trh}
\end{equation}
with $c$ parameterising the contribution from different decay channels. The leading ones are \cite{DR}:
\begin{itemize}
\item \textbf{Higgses}: $c_{\phi \to H_u H_d} = Z^2/12$ where $Z$ controls
the Giudice-Masiero term $K \supset Z\,\frac{H_u H_d}{2\mathcal{V}^{2/3}}\,;$
\item \textbf{Bulk closed string axions}: $c_{\phi \to a_b a_b} =1/24\,;$
\item \textbf{Local closed string axions} (if not eaten by $U(1)$s as in dP cases):
$c_{\phi \to a_{\rm vs} a_{\rm vs}} = 9/384\,.$
\end{itemize}
The strength of the subleading decay channels is instead given by:
\begin{itemize}
\item \textbf{Gauge bosons}: $c_{\phi \to A^{\mu} A^{\mu}} = \lambda \,\alpha_{\rm vs}^2/(8 \pi)\ll 1\,;$
\item \textbf{Other visible sector fields}: $c_{\phi \to \psi \psi} \simeq \left(M_{\rm soft}/m_\phi\right)^2 \simeq 1/\mathcal{V} \ll 1\,;$
\item \textbf{Local open string axions}: $c_{\phi \to a_b \psi_a} \simeq \left(M_s/M_P\right)^4 \tau_{\rm vs}^2
\simeq \left(\tau_{\rm vs}/\mathcal{V}\right)^2 \ll 1\,.$
\end{itemize}

\subsection{Dark radiation}

As can be seen from the leading decay channels above, the branching ratio into light axions
tends to be rather large. The relativistic axions produced in this way behave as DR
since they contribute to the effective number of neutrino-like species $N_{\rm eff}$ defined as:
\begin{equation}
\rho_{\rm rad} = \rho_\gamma \left( 1 + \frac{7}{8}\left( \frac{4}{11} \right)^{4/3} N_{\rm eff} \right),
\end{equation}
where $\rho_{\rm rad}$ is the total radiation energy density whereas $\rho_\gamma$ is the energy density
of all the photons in the universe. $N_{\rm eff}$ is tightly constrained by observations,
$N_{\rm eff}=3.52^{+0.48}_{-0.45}$ at $95\%$ CL \cite{Ade:2013zuv},
which seem to have a slight preference for
an excess of DR at $2\sigma$ with respect to the SM value $N_{\rm eff,SM} = 3.046$:
$\Delta N_{\rm eff}\equiv N_{\rm eff} - N_{\rm eff,SM} \simeq 0.5$.

In the presence of $n_H$ Higgs doublets, $1$ bulk (or volume) axion and $n_a$ local closed string axions,
sequestered LVS models give the following prediction for $\Delta N_{\rm eff}$:
\begin{equation}
\Delta N_{\rm eff}  =  \frac{3.48}{n_H Z^2}\left(1+\frac{9 n_a}{16}\right)
\underset{n_a=0}{\longrightarrow} \frac{3.48}{n_H Z^2}\,.
\label{DNeff}
\end{equation}
Focusing on the case of dP singularities where $n_a=0$, this prediction can give $\Delta N_{\rm eff}\simeq 0.5$
for $Z\simeq 2$ if $n_H=2$ (as in the MSSM)
or for $Z\simeq 1$ if $n_H=6$ (as in some explicit left-right symmetric models \cite{Singular}).
Note however that in the case with a large number of closed string moduli, $n_a \sim \mathcal{O}(100)$
like in a typical axiverse scenario, this prediction yields definitely an overproduction of DR.

\subsection{Axions in sequestered models}
\label{SeqAx}

In LVS models the volume mode is fixed by perturbative $\alpha'$ effects.
Thus the axion $a_b$ remains light because of the shift symmetry. 
Moreover, $a_b$ does not couple to QCD, and so cannot be the QCD axion \cite{Cicoli:2012sz}.
This axion could still be eaten up by an anomalous $U(1)$ living on a bulk cycle.
However, using (\ref{fs}) one has that the axions eaten up are the open string ones since:
$$
K\supset -3\ln\tau_b\quad\Rightarrow\quad\left(f_a^{\rm open}\right)^2 \simeq \left|\frac{\partial K}{\partial \tau_b}\right|=\frac{3}{\tau_b}\gg
\left(f_a^{\rm closed}\right)^2 \simeq \frac{\partial^2 K}{\partial \tau_b^2}=\frac{3}{\tau_b^2}
\quad\text{for}\quad\tau_b\sim\mathcal{V}^{2/3}\gg 1\,.
$$
The final upshot is that ultra-light bulk closed string axions are a \emph{model-independent} feature of LVS models,
and so DR is a generic \emph{prediction} of these string compactifications!

The relativistic axions produced from $\phi$ decay
form a `cosmic axion background' (CAB). They have initially an energy $E_a=m_\phi/2\simeq T_{\rm rh}\sqrt{M_P/m_\phi}\simeq 10^6\,T_{\rm rh}$.
Given that they redshift as photons (up to a small difference since the axions do not thermalise),
this expression can be used to estimate the CAB energy by replacing $T_{\rm rh}$ with the present CMB temperature,
giving an $\mathcal{O}(100)$ eV CAB \cite{Conlon:2013isa}.
These axions have the right energy to account for the observed soft X-ray excess in galaxy clusters
due to their oscillation into photons in the cluster magnetic field~\cite{Conlon:2013txa}.
In order to match the observations one needs and ALP $a_{\rm ALP}$ which is much lighter than the QCD axion $a_{\rm QCD}$
and has an intermediate scale decay constant.

Hence this CAB is populated by at least $a_b$ and $a_{\rm QCD}$, and perhaps $a_{\rm ALP}$ if the observed soft
X-ray excess is due to $a_{\rm ALP}$-$\gamma$ conversion. However in the simplest sequestered models the only light axion is $a_b$. 
Which axions can then behave as $a_{\rm QCD}$ and $a_{\rm ALP}$?
Here are two possibilities:
\begin{itemize}
\item \textbf{Open string QCD axion $\psi_a$}: In this case the axion is the phase of a matter field
$C = |C| \,e^{{\rm i}\psi_a}$ charged under an anomalous $U(1)$. Given that the $\phi$ decay to local open
string axions is subleading, $\psi_a$ gives only a negligible contribution to $\Delta N_{\rm eff}$ without
leading to DR overproduction. From (\ref{fs}) one has:
$$
K\supset \frac{\tau_{\rm vs}^2}{\mathcal{V}}\quad\Rightarrow\quad\left(f_a^{\rm open}\right)^2 \simeq 
\left|\frac{\partial K}{\partial \tau_{\rm vs}}\right|=\frac{2\langle\tau_{\rm vs}\rangle}{\mathcal{V}}\ll
\left(f_a^{\rm closed}\right)^2 \simeq \frac{\partial^2 K}{\partial \tau_{\rm vs}^2}=\frac{2}{\mathcal{V}}
\quad\text{for}\quad\langle\tau_{\rm vs}\rangle\ll 1\,,
$$
implying that the axions eaten up are the closed string ones contrary to geometric regime case 
where they are open string modes.
Subleading F-terms fix $\langle\tau_{\rm vs}\rangle = 1/\mathcal{V}\ll 1$ in the singular regime, 
and so the open string axion decay constant becomes $f_a^{\rm open} \simeq M_s / \sqrt{\mathcal{V}} \simeq \mathcal{O}(10^{11 - 12})\,{\rm GeV}$. 
This is in the phenomenologically allowed window for the QCD axion, avoiding any axionic DM overproduction. 
In this scenario, DM might have two components:
the QCD axion plus Wino/Higgsino non-thermal DM produced from $\phi$ decay~\cite{Allahverdi:2013noa}. 

Sequestered models actually give rise to two light local open string axions 
since any dP singularity yields a gauge theory with two anomalous $U(1)$s. 
A combination of these two light axions could get massive due to QCD instantons and behave as $a_{\rm QCD}$, while the other combination would 
remain massless and play the r\^ole of $a_{\rm ALP}$. Both of these axions would have an intermediate decay constant. 
This scenario leads to interesting predictions:
\begin{enumerate}
\item $a_b$ could account for the observed excess of DR: $\Delta N_{\rm eff}\simeq 0.5$.
\item The QCD axion $a_{\rm QCD}$ could be detected in microwave cavities and $a_{\rm ALP}$ in future light-shining-through-a-wall experiments \cite{Baker:2013zta}.
\item $a_{\rm ALP}$ could explain the transparency of the universe for TeV photons, the anomalous cooling of white dwarfs 
and the soft X-ray excess in galaxy clusters.
\end{enumerate}

\item \textbf{Closed string QCD axion $a_{\rm vs}$}: 
All local closed string axions are eaten by anomalous $U(1)$s in dP singularities but 
some of them could be left over for more complicated singularities. 
The axion decay constant would be set by the string scale 
$f_{a_{\rm vs}}\simeq M_s/\sqrt{4\pi} \simeq 10^{14}$ GeV, leading to axionic DM overproduction 
if $a_{\rm vs}$ is not diluted by $\phi$ decay. However, the reheating temperature in (\ref{Trh}) 
can be rewritten as $T_{\rm rh}\simeq 0.3\,Z$ GeV for $m_\phi \simeq 5\cdot 10^6$ GeV 
which leads to TeV-scale SUSY. In this case the $\phi$ decay to $a_{\rm vs}$ is a leading decay 
channel, and so this axion contributes to DR. The prediction for $\Delta N_{\rm eff}$ 
is given by (\ref{DNeff}) with $n_a=1$: $\Delta N_{\rm eff} \simeq 2.72/Z^2$ (for $n_H=2$). 
$\Delta N_{\rm eff}\simeq 0.5$ can be obtained for $Z\simeq 2.3$, 
implying $T_{\rm rh}\simeq \mathcal{O}(1)$ GeV which is above the QCD phase transition. 
Thus axionic DM cannot be diluted by the $\phi$ decay, and so one has to tune the initial misalignment angle. 
Consequently, this case looks less promising than the one with an open string QCD axion.
\end{itemize}

\section{Acknowledgments}

I would like to thank R. Allahverdi, J. Conlon, B. Dutta, M. Goodsell, 
S. Krippendorf, C. Mayrhofer, F. Quevedo, A. Ringwald, K. Sinha 
and R. Valandro for their collaboration on the topics covered in this review.

\begin{footnotesize}

\end{footnotesize}

\end{document}